\documentclass[conference,a4paper,twocolumn]{IEEEtran}
\usepackage{array}
\newcolumntype{C}[1]{>{\centering\arraybackslash}m{#1}}
\IEEEoverridecommandlockouts

\ifCLASSINFOpdf
\else
\fi
  \usepackage{tipa}
  \usepackage[pdftex]{graphicx}
	\usepackage{epsfig}
	\usepackage{epstopdf}
\usepackage{multirow}

\usepackage[cmex10]{amsmath}
\usepackage{nicematrix}
\usepackage{subfigure}
\usepackage{multirow}

\usepackage{amssymb, amsmath, amsthm, cite}
\usepackage{array}
\usepackage{graphicx}
\usepackage{color}
\usepackage{bbm}
\usepackage{xcolor}

\usepackage{algpseudocode}

\usepackage{algorithm} 
\usepackage{adjustbox}

\newcommand{\beq}[1]{\begin{equation}\label{#1}}
\newcommand{\eeq}{\end{equation}}

\newcommand{\emu}{\end{multline}}
\newcolumntype{P}[1]{>{\centering\arraybackslash}p{#1}}

\newcommand{\A}{\mathcal{A}}

\renewcommand{\H}{\textbf{H}}

\newtheoremstyle{agdTheorem}{\parskip}{\parskip}{\itshape}{\parindent}{\bfseries}{}{0pt}{\thmname{#1}\thmnumber{~#2}.\thmnote{~\textnormal{#3.}}\quad}
\theoremstyle{agdTheorem}

\newtheorem{proposition}{Proposition}

\newtheoremstyle{agdDefinition}{\parskip}{\parskip}{}{\parindent}{\bfseries}{}{0pt}{\thmname{#1}\thmnumber{~#2}.\thmnote{~\textnormal{#3.}}\quad}
\theoremstyle{agdDefinition}

\begin{document}



\title{An Modified Cole's Importance Sampling Method For Low Error Floor  QC-LDPC Codes  Construction}


\author{
\IEEEauthorblockN{
Vasiliy Usatyuk
}
\IEEEauthorblockA{
Department of Information and Computer Technology, South-West State University, Kursk, Russia\\
usa\_uk@ieee.org}
}


%


\maketitle


\begin{abstract}
We modified Cole's Importance Sampling (IS) method for enumerating of Trapping Sets (TS, asymmetric subgraphs) causing an error under message-passing decoder. 
Proposed Cole's IS modifications based on combination of several ideas: parallel TS impulse tree decomposition using unwrapping of message passing iterations, according short cycles dense and straightforward idea of Tanner Graph/ Forney's Normal Graph symmetry - Graph Authomorphism. Its allowed  superior Velasquez-Subramani and Karimi-Banihashemi TS enumerating methods. Particularly proposed  method under PEG (1008, 504) Mackay code for single thread implementation 5027-times (71463 times, multi-treads) faster compare to Velasquez-Subramani LP method and 43-times faster compare to original Cole's method. For TS enumerating problem under (2640, 1320) Margulis code compare to  Velasquez-Subramani LP method proposed method for single thread implementation 37958 times faster, 82-times faster than Karimi-Banihashemi and 134-times faster than Cole's original method. NVIDIA Titan RTX GPU implementation of proposed method gives a further 2-30 times acceleration. 

We show on example of QC-LDPC codes construction how improvement of EMD spectrum, increase hamming(code) distance effect on TS spectrum and BER/FER error-floor level.

\end{abstract}



\begin{IEEEkeywords}
Extrinsic Message Degree (EMD); Importance Sampling (IS); QC-LDPC; Trapping Sets (TS); Dynamic System Theory; Information Geometry; Pseudo-codeword; Convex hull
\end{IEEEkeywords}

%
\IEEEpeerreviewmaketitle


\section{Introduction}

Low-density parity-check (LDPC) codes were first discovered by Gallager~\cite{G65}, generalized  by Tanner~\cite{Tanner81}, Wibberg ~\cite{Wibb96}  and rediscovered by MacKay et al.~\cite{McK96} and Sipser et al.~\cite{SipSpiek96}. 

LDPC codes have been widely used in optical, wireless, satellite, quantum communication and storage systems for providing reliable data store and transmission over different noisy transmission and storage channels. Paradoxically, the recent development of Geometrical Information Theory and its application areas namely the theory of communication and cryptography, has significantly surpassed the development of machine learning, mathematical part theory of: economics, biology, sociology, psychology, military wargaming
 ~\cite{Am,Ri00,ike05,Ma05,Ko06,NiCr17}. 

Due to the possibility to solve with linear complexity the problem of marginalization on graph model (using Belief Propagation method,  Mezard's Cavity method as BP generalization) with relaxing the requirement for accuracy (graph covers as probabilistic relaxation of Graph isomorphism and related positive definite linear model), 
makes LDPC codes and its generalization are the  likely most important part of basis for the organization of Information, Energy and Matter in The Universe. For example, hyperbolic embedding and the hierarchical structure characteristic of statistical manifolds of LDPC codes negative curvature is the main mechanism of how deep learning works, ~\cite{LiTeRo17}. The ability to control systems with delay adequate feedback for making a decision makes the generalization of multigraph LDPC codes on continuous high-dimensional spaces the most likely structures for the emergence of full-fledged artificial intelligence, ~\cite{RU02,Be21}. 

Relaxation of the estimation error due message-passing (Belief propagation) decoder leads to an additional error (complexity  reduction penalty) compared to the maximum likelihood method as a result of cycles presence in the LDPC codes Tanner-graph. Linear size cycles in Tanner graph and their intersections form linear size Trapping Sets (asymmetrical subgraphs) result in a penalty in the waterfall area.  This penalty is calculated using the Covariance evolution method, ~\cite{AmMoRiUr09}. Sublinear size Trapping Sets result in a error-floor penalty from ML decoding. Error-floor penalty estimates using Schlegel-Zhang linear model (~\cite{ScZa10} for BER, bit error rate), generalized (non-integer Hamming weight) union bound (~\cite{Richardson2003ErrorFO,ColeWiHaGi06} for BLER/FER, frame error rate),  and require enumeration of sublinear Trapping sets and weighing them to get sublinear part of generalized weigh spectrum, (non-integer hamming distance due Trapping sets pseudo-codewords) using Importance Sampling techniques. Importance sampling (IS) is a classical statistical (ML) method for probability estimation of rare events,~\cite{Jer84}. In paper ~\cite{Richardson2003ErrorFO} first introduced general approach of trapping sets enumeration and weighing using Monte-Carlo simulation. At papers ~\cite{Fe05,VoKo03} was shown connection of pseudo-codewords of Trapping sets corresponding to the  finite unramified covers of the graph which compromise the message-passing decoder and  fundamental cone in $R^n$ defined by inequalities arising from parity-check matrix $H$ using Feldman linear programming decoding. The search for pseudo-codewords can be carried out using linear programming or IS methods. At papers ~\cite{Cole0Wi,Cole06,Cole08} Cole proposed and applied for LDPC code construction IS enumeration and weighting method. It was one the most well described with application for well know codes and efficient method for TS enumerating. At papers ~\cite{SteChe06,CheSte07,CheSte07_2,CheSte08, CCSV09,CheSte11} Stepanov-Chertkov and Vasic proposed several approaches for trapping set enumeration and weighted for error-floor estimation and found phenomenon of message-passing error due quantization noise. At paper ~\cite{VasCNP09} was given full ontology of Trapping sets for column weight 3 LDPC codes, unfortunately not feasible for codes of greater 3 weight. However, the approach to traversing Tanner-graph cycles in order as the size of the trapping set grows used in many effective Trapping set enumeration methods,  ~\cite{KaBa12,RaDeVa20} including our. At papers ~\cite{AbDeDiRy10,ToBa14,BuSi14,PuSa14,FaMo16,VeSuDra18,PaShiChu12,KiMyJe15,KaBa20} was done LP, IS enumeration of trapping sets and it elimination for LDPC code construction. To compare the efficiency of Trapping sets enumerations, we choose Velazquez-Subramani LP and Cole's, Karimi-Banihashemi, Abu-Surra-Declercq-Divsalar-Ryan IS methods which paper description  contain information about the performance for reference PEG(1008, 504) Mackay, Algebraic (2640, 1320) Margulis LDPC codes was given, (~\cite{AbDeDiRy10,VeSuDra18,ColeWiHaGi06,KaBa12}).

The idea  of  unrolled  iterative belief propagation decoding  algorithms was proposed by Weiberg \cite{Wibb96}.  This idea widely used in many waterfall analysis methods: Density Evolution, Covariance Evolution, Exit-chart, Protograph Exit-chart. This article proposes a further development of this idea for error-floor evaluation. In this paper, a modification of the Cole method is proposed, its efficiency is compared with the methods of Velázquez-Subramani ~\cite{VeSuDra18} and the original Cole method ~\cite{ColeWiHaGi06}. Using the example of constructing an QC-LDPC codes, an lowering error-floor of the LDPC code is demonstrated by increasing the code distance, the EMD spectrum, and it relation to Trapping sets spectrum using the proposed modification of the Cole's IS method.

The main contribution of the paper is a modification of the Cole method, its efficiency is compared with the methods of Velázquez-Subramani ~\cite{VeSuDra18} and the original Cole method ~\cite{ColeWiHaGi06}. Using previous proposed the  Simulated Annealing QC-LDPC code lifting method fifth QC-LDPC codes candidates were constructed, ~\cite{USAVO18}. An improving in the BER/FER, based on low weight pseudocodewords elimination of trapping sets, increasing the code distance, and the improve of EMD spectrum, demonstrated using the proposed modification of the Cole Importance sampling, fast EMD Spectrum estimation. We provide binary or/and source codes of proposed methods: for estimation upper bound on code distance, EMD Spectrum, original and modified Cole's TS enumeration, pseudo-codewords weighing, BER Schelegel-Zhang  linear model.



The outline of the paper is as follows. In Section~\ref{QCLDPC}, we  
introduce some basic definitions such as QC-LDPC codes, Tanner-graph cycle, Trapping Sets, Extrinsic Message Degree metric. In Section~\ref{Coles}, we describe proposed modified Cole's IS method for TS enumeration, performance of 5 QC-LDPC codes obtained by Simulated Annealing lifting method with different code distance and EMD Spectrum and it influence on Trapping Set Spectrum enumerated using the proposed method is investigated by simulations and IS prediction using Schelegel-Zhang linear model (BER), Cole's TS weighted using generalized Union Bound (FER).

 

\section{QC-LDPC Codes: Trapping Sets, EMD Spectrum}\label{QCLDPC}
A QC-LDPC code is described by a parity-check matrix $\H$ which consists of square blocks which could be either zero matrix or circulant permutation matrices.  Let $P=(P_{ij})$ be  the $L\times L$ \textit{circulant permutation matrix} defined by
\beq 
P_{ij}=
\begin{cases}
1,\quad\text{if } i+1\equiv j \mod L\\
0,\quad \text{otherwise}.
\end{cases}
\eeq
Then $P^k$ is the circulant permutation matrix (CPM) which shifts the identity matrix $I$ to the right by $k$ times for any $k$, $0\le k\le L-1$. For simplicity of notation denote the zero matrix by $P^{\infty}$. Denote the set $\{\infty, 0, 1,\ldots, L-1\}$ by $\A_L$. 
Let the matrix $\H$ of size $mL\times nL$ be defined in the following manner
\beq 
 $\H=\left[\begin{array}{cccc} {P^{a_{11} } } & {P^{a_{12} } } & {\cdots } & {P^{a_{1n} } } \\ {P^{a_{21} } } & {P^{a_{22} } } & {\cdots } & {P^{a_{2n} } } \\ {\vdots } & {\vdots } & {\ddots } & {\vdots } \\ {P^{a_{m1} } } & {P^{a_{m2} } } & {\cdots } & {P^{a_{mn} } } \end{array}\right],
\eeq
where $a_{i,j} \in \A_L$. Further we call $L$ the circulant size of $\H$. In what follows a code $C$ with parity-check matrix $\H$ will be referred to as a \textit{QC-LDPC code}.
Let $E(\H) = (E_{ij}(\H))$ be the \textit{exponent matrix} of $\H$ given by:
\beq 
E(\H)=\left[\begin{array}{cccc} {a_{11} } & {a_{12} } & {\cdots } & {a_{1n} } \\ {a_{21} } & {a_{22} } & {\cdots } & {a_{2n} } \\ {\vdots } & {\vdots } & {\ddots } & {\vdots } \\ {a_{m1} } & {a_{m2} } & {\cdots } & {a_{mn} } \end{array}\right],
\eeq
i.e., the entry $E_{ij}(\H) = a_{ij}$. The \textit{mother matrix or base matrix, matrix which graph} $M(\H)$ is a $m\times n$ binary matrix obtained from replacing $-1$'s and other integers by $0$ and $1$, respectively, in $E(\H)$. If there is a cycle of length $2l$ in the Tanner graph of $M(\H)$, it is called a \textit{block-cycle} of length $2l$.  Any block-cycle in $M(\H)$ of length $2l$ corresponds both to the sequence of $2l$ CPM's $\{P^{a_1},P^{a_2},\ldots, P^{a_{2l}}\}$ in $\H$ and sequence of $2l$ integers $\{a_1, a_2, \ldots ,a_{2l}\}$ in $E(\H)$ which will be called \textit{exponent chain}.  The following result gives the fast way to find cycles in the Tanner graph of the matrix $\H$.
\begin{proposition}\textup{\cite{Fossorier04}.}\label{FossCondition}
An exponent chain forms a cycle in the Tanner graph of $\H$ iff the following condition holds

\begin{equation}
\label{eq03}
\sum_{i=1}^{2l}(-1)^i a_i\equiv 0 \mod L.
\end{equation}

\end{proposition}

	Let's consider a sub-graph of the matrix $H$ Tanner graph formed by it's cycles or cycle's overlap. Such a sub-graph includes $a$ variable nodes and $b$ odd degree checks named as trapping set $TS(a,b)$. On the Figure 3 TS(5,3) formed by overlap of three 8-cycles, and TS(4,4) formed by cycle 8 in Tanner graph are presented, ~\cite{VasCNP09}. 
	The hazard of a trapping set depends of the number of variable nodes which could cause the decoding failure on this TS if got errors. 
	For the equal error probability TS(5,3) is more dangerous: with 3 errors in odd degree check nodes it will provide 5 errors in variable nodes while the TS(4,4) produces only 4 errors in variable nodes if there are 4 errors in odd degree check nodes and could be consider as a pseudo codeword of weight 4. The tearing of the most harmful cycles could improve pseudo codeword's weight spectrum and decrease the probability of error-floor. It's hazard is strongly related to the decoder and could be changed by the modification of decoder parameters, ~\cite{Zha-Sch13,CTanJoLi05}.  The minimum codeword that determines the code distance $d_{min}$ of the LDPC code corresponds to the $TS(a, 0)$, $a=d_{min}$,~\cite{VasCNP09}. TS spectrum are a generalization of the weight spectrum of the code in the case of soft iterative decoding by the message-passing decoders. By improving the EMD (ACE), we improve the weight Spectrum  of the LDPC code. Clear that TS spectrum enumeration problem much more complex than weight spectrum enumeration problem. The paper showed the NP complexity of enumerating TS in LDPC codes ~\cite{MCGregMi07}.  

 \begin{figure}
\centering
\includegraphics[width=63mm, viewport=30.00mm 194.40mm 136.75mm 277.00mm]{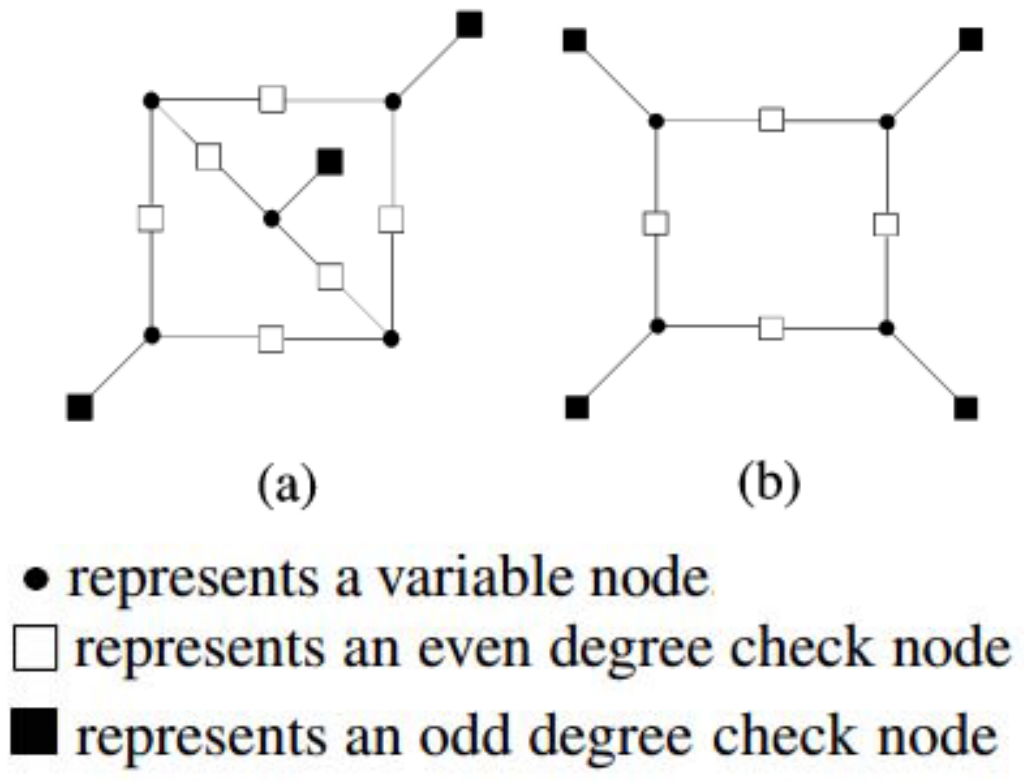} 
  \caption{Graphical representation of  Trapping sets: a) TS(5,3), b) TS(4,4)}
  \label{}
\end{figure}

The metric Extrinsic Message Degree (EMD) of a
cycle in the Tanner graph is defined as the number of
check nodes singly connected to the variable nodes
involved in the cycle. The EMD value of a code is an 
important characteristic, because each cycle is a
trapping set. The EMD metric estimates how strongly
subgraph of cycle is connected with the rest of the
Tanner graph (measure of symmetry). Calculation of the Tanner graph EMD is a hard task, because it requires to determine if the edge is extrinsic edge or cut edge, ~\cite{TiJoViWe}.  We implement  software for EMD Spectrum estimation, source code  available for possibility repeatability of results, ~\cite{EmdSpectrum}.

A LDPC code with a better EMD spectrum for the same code distance (weight enumerator spectrum) under message passsing decoder provides better error-correcting property BER/FER. Such code will not be interfered with by TS-conditioned pseudo-code words. For example,  TS(5,3) has a higher probability of decoding failure than  TS(5,5), and the EMD Spectrum of TS(5,3) will contain 3 cycles of length eight with value 14,  TS(5,5) will contain 3 cycles of length eight with value 16. An increase in the value of the EMD spectrum will lead to a decrease in the probability of TS pseudocodewords decoding failure. 
By construction codes with bigger minimal EMD value we make equal-probability trapping sets elimination. It mean that if TS$(a,b)$ harm depend only from value  $harm =  {b}/{a}$, probability of error in subgraph and multiplicity of this TS. Under this restriction this objective enough to construct codes which not suffer from pseudo-codewords. Unfortunately probability of TS error depend from parameters of decoder: scale and offset values in decoder, scale for message quantization, bit-wise of input bits and message bit, consider error only in information bits, puncturing, shortening and etc, (~\cite{CTanJoLi05}, ~\cite{CCSV09}). By variety this values we can change weight of pseudo-codewords.  
This is reason why precious TS harm model require estimate pseudo-codewords weight using importance sampling and applied union-bound to predict behavior of code in error-floor regime. We use Cole's original TS weighing techniques. This weighed techniques was greatly generalized for different type of channels (AWGN, BEC, BSC and etc) and sizes of TS (sublinear to code length for "error-floor" and linear for waterfall, Covariance Evolution) using local convexity property (topology approach from quantum field theory) of star domain by  paper ~\cite{MiSe19}.

\section{Cole's Importance Sampling modifications}\label{Coles}
\begin{figure}
\includegraphics[width=250 pt]{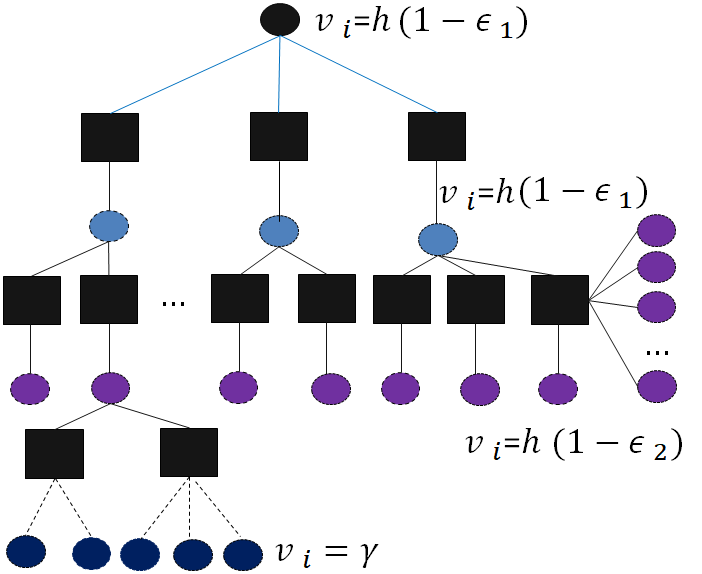}
\caption{\textsf{Cole's Trapping Sets Enumeration Impulse Tree}}
  \label{Fig3}
\end{figure}

\begin{figure*}

\includegraphics[width=\textwidth]{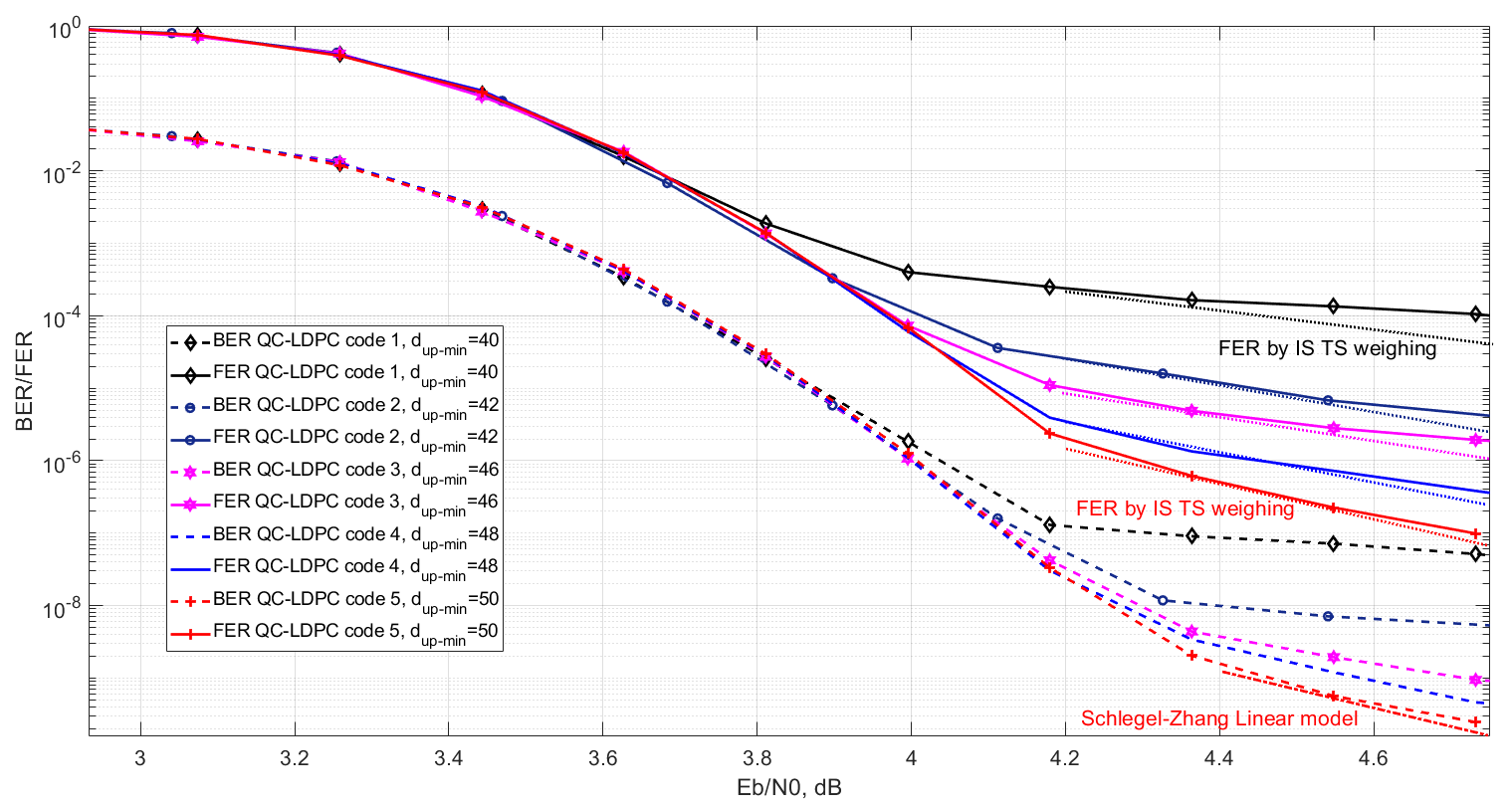}
\caption{\textsf{Simulation results using Quantized Normalized Layer Min-Sum decoder under BPSK and IS estimation of BER (Linear Model based), FER (TS weighted Union Bound based) error-floor for 5 QC-LDPC codes $4\times 20$ (2560, 2048) constructed with different values of EMD Spectrum, $d_{min}$}}
  \label{Fig2}
\end{figure*}

\begin{table*}[!htb]
\caption{Execution time (in seconds) expended to solve TS problem enumerating (in
seconds) is reported for the \newline PEG(1008, 504) Mackay code and Algebraic(2640, 1320) Margulis code.}
\label{tab01}
\center
\begin{tabular}{|C{0.5in}|C{0.4in}|C{1.in}|C{0.6in}|C{0.4in}|C{0.5in}|C{0.6in}|C{0.6in}|C{0.6in}|} 

\hline 
Code & \multicolumn{3}{|C{1.9in}|}{PEG(1008, 504) Mackay code} & \multicolumn{5}{|C{3.2in}|}{Algebraic (2640, 1320) Margulis code} \\ 
\hline 
 $a$ value of  $TS(a, b)$ & LP ~\cite{VeSuDra18} & Proposed method &
Original\newline Cole's, ~\cite{ColeWiHaGi06} & LP ~\cite{VeSuDra18} & Proposed method   & Modified\newline Cole's, ~\cite{KaBa12} & Modified \newline Cole's, ~\cite{AbDeDiRy10} & Original\newline Cole's, ~\cite{ColeWiHaGi06} \\ \hline 


 5& 53.42 &\multirow{8}{*}{99.5,  7 (multitreads)}  & \multirow{8}{*}{4320  (1.2h)} & 520.95 & \multirow{5}{*}{0.97} & \multirow{10}{*}{\parbox{1.cm}{18000 (5h)}} & \multirow{10}{*}{\parbox{1.cm}{ 604800 \newline  (7 days) }}&\multirow{10}{*}{\parbox{1.cm}{  29520 \newline (8.2h)}} \\ \cline{1-2} \cline{5-5}

6 & 31.74 &{}  &  & 4373.78 &  &  &  &  \\ \cline{1-2} \cline{5-5}
7 & 984.51 & {}  &  & 716.04 &  &  &  &  \\ \cline{1-2} \cline{5-5}
8 & 4.1 &  &   & 18843.19 &  &  &  &  \\ \cline{1-2} \cline{5-5}
9 & 4864.37 &   &  & 13504.13 &  &  &  &  \\ \cline{1-2} \cline{5-6}
10 & 8143.11 &   &  &  & \multirow{5}{*}{\parbox{1cm}{3479, 219 (multitreads)}}&  &  &  \\ \cline{1-2} \cline{5-5}
11 & 485919.5 &   &  &  &  &  &  &  \\ \cline{1-2} \cline{5-5} 
12 & 245.1 &   &  &  &  &  &  &  \\  \cline{1-5}
\dots  &  &  &  &  &  &  &  &  \\ \cline{1-5}
18 &  &  &  &  &  &  &  &  \\ \hline 
\end{tabular}

\end{table*}

To enumerate Trapping Sets in  arbitrary LDPC codes Cole proposed  unrolled  iterative belief propagation computation tree in Density evolution like manner to apply modified Tree message-passing deterministic noise error impulses method previously used to enumerated code distance proposed by Berrou ~\cite{BeVa} and Hu-Fossorier-Eleftheriou  ~\cite{HuFoEl}.  The possibility of using this method is due to the previously demonstrated fact that trapping sets generalize the weight spectrum of codes.  

For each variable node ${\rm VN}_{{\rm i}}^{{\rm 1}} $ of degree not larger than $\lambda _{\max }$, a tree of depth 3 (${\rm VN}^{{\rm 1}} {\rm -CN}^{{\rm 1}} {\rm -VN}^{{\rm 2}} {\rm -CN}^{{\rm 2}} {\rm -VN}^{{\rm 3}} {\rm -CN}^{{\rm 3}} $) or larger is built, Fig. ~\ref{Fig3}. For each set of variable nodes ${\rm VN}^{{\rm i}} $, an error impulse is generated by the following rule: for all variable nodes from the set, the log-likelihood value is initialized with $h\left(1-\varepsilon _{i} \right)$ and last layers are set to $\varepsilon _{i} h$, where $\varepsilon _{i} $- magnitude of error impulse,  $h$- channel variance. For example, under AWGN-channel BPSK $\frac{2}{\sigma ^{2} } $, $\gamma \in (0,1]$. 

Lets $w_H\left(Hx_{iter}\right)$ is number of unsatisfied parity checks after $iter$ message passing iteration, $H$ -- LDPC parity-check matrix, $x_{iter} $- information word.  Subgraph induced by set $S\left(\min _{iter} \omega _{H} (Hx_{iter} \right)$ is potential Trapping set for check according to definition.  To further improve Cole's method we modified original sets $S_{1}, S_{2}$ at Step 1, ~\cite{ColeWiHaGi06}. We introduce Cole's method major modification -  subgraph separation function based on \textbf{mandatory requirement of subgraph participation in the cycle}, according to trapping sets definition. First, for a given set  ${\rm VN}$ of variable nodes, define $TS({\rm VN}{\rm )}$ as the union of ${\rm VN}$ and all variable nodes which add a new cycle to the subgraph induced by ${\rm VN}$. Second, define ${\rm Tree(VN)}$ as the most powerful subset of ${\rm VN}$ such that the subgraph induced by it doesn't have any tree-like structures. Using these two sets, let define harmful ordered tree subgraph sieving sequences:
\begin{center}
$
 {S_{1} (x)=Tree(S(x)\bigcup \{ VN_{1} ,VN_{2} ...,VN_{\lambda } \} ),}$ \\
 ${S_{2} (x)=Tree(TS(S(x)\bigcup \{ VN_{1} ,VN_{2} ...,VN_{\lambda } \} )),} 
 $
\end{center}

where $\{ VN_{1} ,VN_{2} ...,VN_{\lambda } \} $ is the set of variable nodes with error impulse. For a given set ${\rm VN}$ of variable nodes in subgraph, TS (a,b), a=  $\left|{\rm VN}\right|$ and b odd-degree check nodes in the subgraph formed by ${\rm VN}$. For trapping sets easy to define ordering rule from $b/a$ ration: ${\rm TS(a,b)}\prec {\rm TS(a',b'):b=b'\& a}\le {\rm a'.}$ We can get ordered set: 
\begin{center}
$ S_{1} (x_{\left\lfloor girth/2\right\rfloor } ),S_{2} (x_{\left\lfloor girth/2\right\rfloor } ),...,S_{1} (x_{\max } ),S_{2} (x_{\max } )$
\end{center}

by ordering rule we can take ${\rm VN}$ set from list, if it is not empty, we consider the subgraph induced by it as a trapping set. Such ordered set could be easily decomposed to increase parallelism level. Using average number of iteration to converge at require SNR level, we can make decoding by our used version of message passing decoder If the impulse was successfully decoded we choose another error impulses. If the impulse was not decoded we do the remaining iteration to check. After each iteration step we form $S_{1} (x_{\lceil avg \rceil} ),S_{2} (x_{\lceil avg \rceil} ),...,S_{1} (x_{\max } ),S_{2} (x_{\max } )$ sets. Such approach allow to make several stage scheduler for processing and control parallelism level, RAM consumption according hardware limitation and varying the number of iterations according to the density of cycles in the graph passing through the symbol node. In TS formed by short cycles, a fatal decoding error occurs with a smaller number of iterations. Fast calculation of the cycle density in a Tanner graph can be done by the method described at paper  ~\cite{HaCh06}, implementation  ~\cite{HaCh06Im} and represented at Table V. 
Comparison of the TS enumeration time for the proposed modified Cole's method, LP method and IS methods is given in the Table I.
Proposed  method under (1008, 504) LDPC code for single thread implementation 5027-times faster compare to LP method and 43-times faster compare to original Cole's method. For TS enumerating problem under (2640, 1320) LDPC code compare to LP method proposed method for single thread implementation 37958 times faster, 82-times faster than Karimi-Banihashemi and 134-times faster than Cole's original method.  We lift using Simulated annealing method (~\cite{USAVO18}) 5 QC-LDPC codes from same mother matrix (Table V), with different EMD Spectrum and code distance (Table II), to show how EMD Spectrum and code distance affect on Trapping Sets (Table III, IV).  Memory consumption, for 4x20 QC-LDPC codes (Table V) from example contain 77 non-zero circulants, require maximal memory around 16 Gb and allow to use small iteration buffer at CPU after make arbitrary tread occupation for NVIDIA TESLA GPU or FPGA accelerator board. Modern FPGA Accelerator Board like BittWare XUP-P3R FPGA can be equipped with large high speed RAM and allow to make large part of multi stage processing on FPGA device providing further acceleration from 1.25 (small size of protograph and high code rates) to 44  times (large than 200 variable nodes in protograph and small code rate). The simulation results show how the improvement of the EMD spectrum and increasing the code distance in the simulated annealing lifting method, ~\cite{USAVO18}, leads to error-floor level lowering, Fig. ~\ref{Fig2}. The simulation done using  20 iterations Quantized message(LLRs=4, C2V=4, V2C=6 bits) Normalized Layer Min-Sum decoder, continued until the accumulation of 200 frame errors. 
 I want to thank Dr. Chad A. Cole for providing the source code of the IS method. 



\begin{table*}[t]

\caption{EMD Spectrum, without multiplication by graph automorphism 128, for 5 QC-LDPC codes with same base matrix}
\label{tab01}
\center
\begin{tabular}{|c|c|c|c|c|c|} \hline 
Cycle size, EMD & Code 1,  $d_{upper}=40$ & Code 2,  $d_{upper}=42$ & Code 3,  $d_{upper}=46$ & Code 4,   $d_{upper}=48$ & Code 5,   ${\ d}_{upper}=50$ \\ \hline 
 & \multicolumn{5}{|c|}{Number of cycles with } \\ \hline 
4,2 & 1 & 0 & 0 & 0 & 0 \\ \hline 
4,3 & 0 & 2 & 1 & 0 & 0 \\ \hline 
4,4 & 0 & 0 & 5 & 0 & 0 \\ \hline 
6,2 & 2 & 0 & 0 & 0 & 0 \\ \hline 
6,3 & 0 & 2 & 0 & 0 & 0 \\ \hline 
6,4 & 1 & 1 & 7 & 5 & 0 \\ \hline 
6,5 & 35 & 36 & 39 & 45 & 47 \\ \hline 
6,6 & 125 & 143 & 118 & 113 & 115 \\ \hline 
8,2 & 1 & 0 & 0 & 0 & 0 \\ \hline 
8,3 & 2 & 1 & 1 & 0 & 0 \\ \hline 
8,4 & 9 & 3 & 23 & 1 & 2 \\ \hline 
8,5 & 28 & 92 & 88 & 29 & 41 \\ \hline 
8,6 & 242 & 344 & 492 & 263 & 260 \\ \hline 
8,7 & 2122 & 2050 & 2134 & 2160 & 2161 \\ \hline 
8,8 & 5381 & 5227 & 4991 & 5302 & 5494 \\ \hline 
10,2 & 2 & 0 & 0 & 0 & 0 \\ \hline 
10,3 & 2 & 2 & 0 & 0 & 0 \\ \hline 
10,4 & 18 & 12 & 31 & 4 & 2 \\ \hline 
10,5 & 218 & 186 & 203 & 41 & 43 \\ \hline 
10,6 & 962 & 1160 & 1371 & 501 & 418 \\ \hline 
10,7 & 4049 & 5802 & 7653 & 3773 & 3761 \\ \hline 
10,8 & 20806 & 24402 & 31326 & 21476 & 21896 \\ \hline 
10,9 & 96694 & 93137 & 97284 & 96047 & 95060 \\ \hline 
10,10 & 187719 & 185905 & 172856 & 189317 & 188391 \\ \hline 
12,2 & 3 & 0 & 0 & 0 & 0 \\ \hline 
12,3 & 15 & 3 & 5 & 0 & 0 \\ \hline 
12,4 & 62 & 39 & 120 & 10 & 5 \\ \hline 
12,5\textbf{} & 461 & 575 & 690 & 153 & 175 \\ \hline 
12,6 & 3411 & 3728 & 5214 & 1326 & 1153 \\ \hline 
12,7 & 20275 & 21277 & 29596 & 10123 & 9447 \\ \hline 
12,8 & 88824 & 113108 & 150890 & 69143 & 65034 \\ \hline 
12,9 & 373262 & 469656 & 610107 & 365459 & 369820 \\ \hline 
12,10 & 1473818 & 1604858 & 1922131 & 1498704 & 1535790 \\ \hline 
12,11 & 4660137 & 4464438 & 4550663 & 4611680 & 4614646 \\ \hline 
12,12 & 7218738 & 7154661 & 6563134 & 7284713 & 7253616 \\ \hline 
14,2 & 2 & 0 & 0 & 0 & 0 \\ \hline 
14,3 & 30 & 6 & 4 & 4 & 0 \\ \hline 
14,4 & 222 & 126 & 253 & 26 & 21 \\ \hline 
14,5 & 2028 & 1525 & 2241 & 363 & 404 \\ \hline 
14,6 & 13536 & 12887 & 18045 & 3629 & 3650 \\ \hline 
14,7 & 74009 & 83587 & 120117 & 32418 & 31056 \\ \hline 
14,8 & 416600 & 472535 & 665918 & 234832 & 223210 \\ \hline 
14,9 & 2011841 & 2306984 & 3159570 & 1406425 & 1349521 \\ \hline 
14,10 & 8006875 & 9752033 & 12759048 & 7035658 & 6935280 \\ \hline 

\end{tabular}

\end{table*}

\begin{table*}
\caption{Ts(a,b) enumeration for Code 1, $b<19$ from 29,  $d_{upper}=40$, $Girth=4$, $EMD=2$,  full list available at~\cite{ColeTS}}
\begin{tabular}{|c|c|c|c|c|c|c|c|c|c|c|c|c|c|c|c|c|c|} \hline 
 & b=2 & 3 & 4 & 5 & 6 & 7 & 8 & 9 & 10 & 11 & 12 & 13 & 14 & 15 & 16 & 17 & b=18 \\ \hline 
a=2 & 128 & 0 & 0 & 0 & 0 & 0 & 0 & 0 & 0 & 0 & 0 & 0 & 0 & 0 & 0 & 0 & 0 \\ \hline 
3 & 128 & 0 & 0 & 0 & 0 & 0 & 0 & 0 & 0 & 0 & 0 & 0 & 0 & 0 & 0 & 0 & 0 \\ \hline 
4 & 0 & 0 & 520 & 0 & 0 & 0 & 0 & 0 & 0 & 0 & 0 & 0 & 0 & 0 & 0 & 0 & 0 \\ \hline 
5 & 0 & 6 & 344 & 393 & 0 & 0 & 0 & 0 & 0 & 0 & 0 & 0 & 0 & 0 & 0 & 0 & 0 \\ \hline 
6 & 0 & 3 & 196 & 671 & 3728 & 0 & 0 & 0 & 0 & 0 & 0 & 0 & 0 & 0 & 0 & 0 & 0 \\ \hline 
7 & 0 & 5 & 105 & 744 & 5109 & 12434 & 0 & 0 & 0 & 0 & 0 & 0 & 0 & 0 & 0 & 0 & 0 \\ \hline 
8 & 0 & 0 & 98 & 915 & 6959 & 15427 & 22615 & 0 & 0 & 0 & 0 & 0 & 0 & 0 & 0 & 0 & 0 \\ \hline 
9 & 0 & 2 & 130 & 888 & 5102 & 18550 & 24858 & 29106 & 0 & 0 & 0 & 0 & 0 & 0 & 0 & 0 & 0 \\ \hline 
10 & 0 & 2 & 113 & 677 & 3492 & 11791 & 30707 & 30834 & 30306 & 0 & 0 & 0 & 0 & 0 & 0 & 0 & 0 \\ \hline 
11 & 0 & 3 & 102 & 510 & 2303 & 7106 & 16947 & 31194 & 28352 & 24092 & 0 & 0 & 0 & 0 & 0 & 0 & 0 \\ \hline 
12 & 0 & 14 & 73 & 383 & 1482 & 4125 & 9075 & 17218 & 26066 & 23090 & 19612 & 0 & 0 & 0 & 0 & 0 & 0 \\ \hline 
13 & 0 & 8 & 58 & 243 & 849 & 2194 & 4779 & 9119 & 15040 & 20367 & 18421 & 14976 & 0 & 0 & 0 & 0 & 0 \\ \hline 
14 & 1 & 7 & 29 & 165 & 519 & 1184 & 2360 & 4735 & 8163 & 12143 & 15642 & 14174 & 11477 & 0 & 0 & 0 & 0 \\ \hline 
15 & 0 & 3 & 18 & 77 & 260 & 575 & 1195 & 2388 & 4132 & 6667 & 9387 & 11831 & 11049 & 8728 & 0 & 0 & 0 \\ \hline 
16 & 0 & 2 & 8 & 40 & 112 & 245 & 519 & 1059 & 1969 & 3369 & 5247 & 7252 & 8871 & 8333 & 6692 & 0 & 0 \\ \hline 
17 & 0 & 1 & 2 & 13 & 53 & 111 & 252 & 482 & 907 & 1634 & 2767 & 3970 & 5494 & 6479 & 6322 & 5270 & 0 \\ \hline 
18 & 0 & 0 & 2 & 2 & 16 & 48 & 131 & 223 & 413 & 759 & 1387 & 2163 & 3239 & 4168 & 5007 & 4827 & 3943 \\ \hline 
19 & 0 & 0 & 2 & 2 & 4 & 21 & 53 & 103 & 164 & 325 & 656 & 1069 & 1660 & 2421 & 3200 & 3952 & 3925 \\ \hline 
20 & 0 & 0 & 0 & 1 & 2 & 14 & 19 & 44 & 85 & 126 & 259 & 548 & 856 & 1379 & 2010 & 2577 & 2904 \\ \hline 
21 & 0 & 2 & 0 & 0 & 2 & 8 & 5 & 15 & 39 & 85 & 151 & 250 & 393 & 723 & 1084 & 1641 & 2090 \\ \hline 
22 & 0 & 0 & 0 & 0 & 1 & 3 & 4 & 9 & 18 & 31 & 54 & 112 & 220 & 359 & 579 & 950 & 1339 \\ \hline 
23 & 0 & 0 & 0 & 0 & 0 & 0 & 1 & 1 & 5 & 12 & 23 & 52 & 94 & 168 & 298 & 518 & 736 \\ \hline 
24 & 0 & 0 & 0 & 0 & 0 & 0 & 0 & 1 & 2 & 3 & 12 & 27 & 40 & 109 & 160 & 251 & 447 \\ \hline 
25 & 0 & 0 & 0 & 0 & 0 & 0 & 0 & 0 & 2 & 1 & 6 & 7 & 8 & 39 & 64 & 131 & 211 \\ \hline 
26 & 0 & 0 & 0 & 0 & 0 & 0 & 0 & 0 & 0 & 0 & 2 & 1 & 7 & 12 & 35 & 43 & 102 \\ \hline 
27 & 0 & 0 & 0 & 0 & 0 & 0 & 0 & 0 & 0 & 0 & 0 & 0 & 0 & 3 & 9 & 28 & 50 \\ \hline 
28 & 0 & 0 & 0 & 0 & 0 & 0 & 0 & 0 & 0 & 0 & 0 & 0 & 0 & 1 & 1 & 9 & 24 \\ \hline 
a=29 & 0 & 0 & 0 & 0 & 0 & 0 & 0 & 0 & 0 & 0 & 0 & 0 & 0 & 0 & 2 & 4 & 1 \\ \hline

\end{tabular}

\end{table*}

\begin{table*}
\caption{Ts(a,b) enumeration for Code 5, $b<19$  from 29,  $d_{upper}=50$, $Girth=6$, $EMD=5$,  full list available at~\cite{ColeTS}}
\begin{tabular}{|c|c|c|c|c|c|c|c|c|c|c|c|c|c|c|c|c|c|} \hline 
 & b=2 & 3 & 4 & 5 & 6 & 7 & 8 & 9 & 10 & 11 & 12 & 13 & 14 & 15 & 16 & 17 & b=18 \\ \hline 
a=2 & 0 & 0 & 0 & 0 & 0 & 0 & 0 & 0 & 0 & 0 & 0 & 0 & 0 & 0 & 0 & 0 & 0 \\ \hline 
3 & 0 & 0 & 0 & 0 & 0 & 0 & 0 & 0 & 0 & 0 & 0 & 0 & 0 & 0 & 0 & 0 & 0 \\ \hline 
4 & 0 & 0 & 79 & 0 & 0 & 0 & 0 & 0 & 0 & 0 & 0 & 0 & 0 & 0 & 0 & 0 & 0 \\ \hline 
5 & 0 & 0 & 5 & 174 & 0 & 0 & 0 & 0 & 0 & 0 & 0 & 0 & 0 & 0 & 0 & 0 & 0 \\ \hline 
6 & 0 & 0 & 12 & 267 & 2700 & 0 & 0 & 0 & 0 & 0 & 0 & 0 & 0 & 0 & 0 & 0 & 0 \\ \hline 
7 & 0 & 0 & 35 & 323 & 3218 & 14341 & 0 & 0 & 0 & 0 & 0 & 0 & 0 & 0 & 0 & 0 & 0 \\ \hline 
8 & 0 & 0 & 46 & 337 & 4302 & 14497 & 31041 & 0 & 0 & 0 & 0 & 0 & 0 & 0 & 0 & 0 & 0 \\ \hline 
9 & 0 & 0 & 14 & 467 & 3481 & 16503 & 24757 & 37520 & 0 & 0 & 0 & 0 & 0 & 0 & 0 & 0 & 0 \\ \hline 
10 & 0 & 0 & 21 & 451 & 2526 & 9403 & 28995 & 32280 & 39235 & 0 & 0 & 0 & 0 & 0 & 0 & 0 & 0 \\ \hline 
11 & 0 & 0 & 34 & 350 & 1804 & 5786 & 15297 & 30175 & 28766 & 27256 & 0 & 0 & 0 & 0 & 0 & 0 & 0 \\ \hline 
12 & 0 & 0 & 20 & 234 & 1135 & 3417 & 7902 & 14947 & 22918 & 21173 & 19536 & 0 & 0 & 0 & 0 & 0 & 0 \\ \hline 
13 & 0 & 0 & 24 & 164 & 673 & 1743 & 3738 & 7014 & 11793 & 16750 & 15667 & 14445 & 0 & 0 & 0 & 0 & 0 \\ \hline 
14 & 0 & 2 & 19 & 81 & 326 & 841 & 1590 & 3131 & 5624 & 8955 & 11895 & 11079 & 10393 & 0 & 0 & 0 & 0 \\ \hline 
15 & 0 & 0 & 8 & 38 & 137 & 345 & 643 & 1340 & 2374 & 4082 & 6310 & 8504 & 8209 & 7542 & 0 & 0 & 0 \\ \hline 
16 & 0 & 0 & 2 & 15 & 65 & 125 & 274 & 446 & 979 & 1821 & 2993 & 4515 & 6183 & 6124 & 5573 & 0 & 0 \\ \hline 
17 & 0 & 0 & 2 & 3 & 21 & 39 & 85 & 185 & 332 & 775 & 1348 & 2219 & 3296 & 4425 & 4467 & 4064 & 0 \\ \hline 
18 & 0 & 0 & 0 & 1 & 7 & 8 & 31 & 73 & 161 & 269 & 545 & 984 & 1632 & 2446 & 3206 & 3312 & 2976 \\ \hline 
19 & 0 & 0 & 0 & 0 & 1 & 8 & 10 & 28 & 46 & 116 & 222 & 443 & 762 & 1230 & 1894 & 2458 & 2640 \\ \hline 
20 & 0 & 0 & 0 & 0 & 1 & 0 & 2 & 9 & 20 & 41 & 90 & 186 & 334 & 618 & 995 & 1467 & 1860 \\ \hline 
21 & 0 & 0 & 0 & 0 & 0 & 0 & 1 & 1 & 7 & 19 & 40 & 80 & 160 & 298 & 515 & 803 & 1240 \\ \hline 
22 & 0 & 0 & 0 & 0 & 0 & 0 & 0 & 1 & 3 & 8 & 18 & 26 & 68 & 126 & 253 & 398 & 687 \\ \hline 
23 & 0 & 0 & 0 & 0 & 0 & 0 & 0 & 1 & 1 & 3 & 5 & 6 & 27 & 64 & 115 & 194 & 363 \\ \hline 
24 & 0 & 0 & 0 & 0 & 0 & 0 & 1 & 0 & 0 & 0 & 3 & 8 & 15 & 22 & 59 & 97 & 153 \\ \hline 
25 & 0 & 0 & 0 & 0 & 0 & 0 & 0 & 0 & 0 & 1 & 0 & 3 & 1 & 5 & 17 & 44 & 93 \\ \hline 
26 & 0 & 0 & 0 & 0 & 0 & 0 & 0 & 0 & 0 & 0 & 0 & 0 & 3 & 2 & 7 & 15 & 28 \\ \hline 
27 & 0 & 0 & 0 & 0 & 0 & 0 & 0 & 0 & 0 & 0 & 0 & 0 & 0 & 1 & 3 & 7 & 13 \\ \hline 
28 & 0 & 0 & 0 & 0 & 0 & 0 & 0 & 0 & 0 & 0 & 0 & 0 & 0 & 0 & 0 & 3 & 4 \\ \hline 
a=29 & 0 & 0 & 0 & 0 & 0 & 0 & 0 & 0 & 0 & 0 & 0 & 0 & 0 & 0 & 0 & 0 & 1 \\ \hline 

\end{tabular}

\end{table*}

\begin{table*}

\caption{Code 1 QC Parity-check matrix and average number of cycles 4, 6 passing through the row, column related to 128}
\begin{tabular}{|c|c|c|c|c|c|c|c|c|c|c|c|c|c|c|c|c|c|c|c|c|c|} \hline 

Cyc6 & ~ & 3,67 & 5 & 5 & 7& 8,6 & 7,6 & 11& 11 & 7 & 9 & 7,6 & 6,6 & 9,3 & 9,3 & 7,3 & 9,3 & 9,67 & 9 & 9 & 10,6 \\ \hline  
~ & Cyc4 & 0,5 & 0 & 0,5 & 0 & 0 & 0 & 0 & 0 & 0 & 0 & 0 & 0 & 0 & 0 & 0 & 0 & 0 & 0 & 0 & 0 \\ \hline 
39,67 & 0,5 & 34 & 1 & 81 & 20 & 0 & 0 & 0 & 8 & 54 & 0 & 44 & 111 & 28 & 99 & 75 & 81 & 81 & 42 & 30 & 119 \\ \hline 
38,33 & 0 & -1 & 85 & 36 & 18 & 113 & 59 & 29 & 58 & 123 & 38 & 3 & 91 & 113 & 1 & 0 & 89 & 72 & 44 & 78 & 90 \\ \hline 
42,33 & 0,5 & 122 & -1 & 41 & 71 & 10 & 85 & 110 & 66 & 99 & 77 & 109 & 100 & 0 & 43 & 33 & 124 & 26 & 98 & 74 & 28 \\ \hline 
42,67 & 0 & 109 & 103 & -1 & 75 & 36 & 12 & 58 & 127 & 53 & 114 & 1 & 18 & 48 & 116 & 120 & 15 & 51 & 63 & 95 & 68 \\ \hline 
\end{tabular}

\end{table*}


\section{Conclusion}
  The proposed modified Cole’s IS
method for enumerating of Trapping Sets significantly outperforms previously proposed methods in terms of TS enumeration speed and parallel processing capabilities due to the unwrapping of the iteration tree and use of graph automorphism. The example of 5 QC-LDPC codes construction demonstrates how simulated annealing lifting with code candidates EMD spectrum improving, hamming distance sieving eliminate harmfully Trapping sets which leads to a lowering of the error-floor level.





%


\end{document}